# A labeled Clinical-MRI dataset of Nigerian brains

## Authors
Eberechi Wogu[1*], Patrick Filima[1*], Bradley Caron[2], Daniel Levitas[2], Peer Herholz[2], Catherine Leal[2], Mohammed F. Mehboob[2], Soichi Hayashi[2], Simisola Akintoye[3], George Ogoh[4], Tawe Godwin[5], Damian Eke[3], Franco Pestilli[2]

## Affiliations
1. University of Port Harcourt, Choba, Rivers State, Nigeria.
2. The University of Texas at Austin, TX USA
3. Center for Law, Justice and Society, De Montfort University, UK.
4. Center for Computing and Social Responsibility, De Montfort University, UK.
5. Lifebridge medical diagnostic Center, Garki 2, Abuja Nigeria.

* These author contributed equally to this work

## Abstract
We describe a Magnetic Resonance Imaging (MRI) dataset from individuals from the African nation of Nigeria. The dataset contains pseudonymized structural MRI (T1w, T2w, FLAIR) data of clinical quality. Dataset contains data from 36 images from healthy control subjects, 32 images from individuals diagnosed with age-related dementia and 20 from individuals with Parkinson's disease. There is currently a paucity of data from the African continent. Given the potential for Africa to contribute to the global neuroscience community, this first MRI dataset represents both an opportunity and benchmark for future studies to share data from the African continent.

## Background & Summary

Human populations in Africa have among the greatest genetic diversity in the world. Mental and brain health are also among the most diverse in the world, because of the variety of stress factors, such as child health, the high level of incidence of traumatic brain injury and the diseases endemic to the region[1]. Concurrently, Africa is experiencing a demographic shift with rising life expectancy, resulting in an increasing population of individuals aged 60 or above. For example, in Nigeria approximately 7.76 million citizens out of a total population of 230 millions (3.36%) are above 65 years old and the number is increasing[2–4]. This trend is accompanied by an increased prevalence in age-related disease[2], such as Parkinson's Disease (PD)[5] and Dementia[6]. In contrast with the needs and opportunities offered by Africa, Africa's neuroscientific research is far from realizing its full potential[1]. Opportunities to increase Africa's neuroscientific output and to study African brains are currently limited because of the paucity of available data on the African populations.

Here we contribute to advancing brain understanding by describing the first openly-shared Magnetic Resonance Imaging (MRI) dataset from individuals from the African nation of Nigeria. Study participants were recruited in three diagnostic centers across two geopolitical regions in Nigeria; the South-South region and the North-Central region spanning across the following states: Delta, Edo, Bayelsa, Cross River, Akwa Ibom and Rivers States (for the South-South) and Benue, Kogi, Kwara, Nasarawa, Niger, Plateau States and the Federal Capital Territory (for the North-Central). These data are unique because of the paucity of available shared data in similar ethnic groups. However, due to the lower strength of the MRI scanners used in procurement of the data (0.3 and 1.5 Tesla), the data is marked by diminished quality as they were collected following the best clinical practices in the regions and diagnostic centers without research purposes nor protocol standardization. The diversity in the ethnic group, the data quality, and protocol heterogeneity, provide a first opportunity for the training of local clinicians, radiologists and researchers. Furthermore, the dataset provides a unique opportunity for medical imaging and artificial intelligence systems interested in developing algorithms for the detection of disease that are robust to diminished data quality and heterogeneity.

Parkinson's disease (PD) is a gradually progressive neurodegenerative movement disorder resulting from selective loss of nigral dopaminergic neurons of unknown etiology, and is clinically characterized by both motor and non-motor manifestations[5]. Age-related dementia, on the other hand, is a syndrome characterized by deterioration in cognitive function beyond what might be expected from the usual consequences of biological aging. It is caused by damage to or loss of nerve cells and their connections in the brain regions associated with memory and learning[6]. Parkinson's Disease (PD) and Dementia are most prevalent among people 60 and older, and are consequently among the leading causes of disability and dependency among older people globally[2]. PD has a prevalence rate of 6.0% to 8.3% of neurologic hospital admissions/consultations in West Africa, with estimated crude prevalence varying from 15 to 572 per 100,000 people[7,8]. Dementia, on the other hand, has a prevalence rate of 2.3% to 20.0%, and the incidence rates are 13.3 per 1000 people with increased mortality in parts of rapidly developing Africa[6]. Despite the rising prevalence of PD and Dementia in Africa, there is relatively little information on them in the global brain data repositories [5,9].

Diversity or heterogeneity of datasets is critical to an effective, functional, dependable and trustworthy health ecosystem. However, datasets from the global south, particularly from Africa, are worryingly missing from the global neuroscience research and innovation space. Furthermore, the availability of Findable, Accessible, Interoperable and Reusable (FAIR) data[10] is a critical factor that drives global health research and innovation. Despite comprising 12.5% of the world's population, Africa still accounts for less than 1 percent of global research output[11]. Adequate representation of diverse datasets that reflects global demographics in scientific research is crucial for inclusion and equitable global health research. Our research aims to make FAIR African brain data available using Nigerian brain MRI datasets of patients with Parkinson's Disease and Dementia as use cases.

We share the data utilizing a recently developed unique approach that exploits the free and secure cloud computing platform brainlife.io. These datasets were converted from DICOM to BIDS using ezBIDS (https://brainlife.io/ezbids), which then transferred the data to brainlife.io. This approach integrates, into a single record both the data and reproducible web-services implementing the full processing pipeline[12–14].

The processed data contains derivatives across 93 participants (Dementia: 32; Parkinson: 19; Control: 42) (**see Table 1 and Table 2 for data sources and demographics, respectively**). The total size of the repository is approximately 0.34212 GB of imaging data and derivatives, comprising T1w, T2w, and FLAIR anatomicals, brain masks, and summary measures reported from MRIQC. The processing pipeline implemented to process this dataset utilizes mainstream neuroimaging software libraries including FSL[15–17] and MRIQC[18]. The corresponding brainlife.io apps were developed with a lightweight specification and use modern methods for software containerization making the analyses trackable, reproducible, and reusable on a wide range of computing resources[19]. The present descriptor describes the repository and pipelines published via brainlife.io mechanisms. These resources will allow the broader research community to gain additional insight into the pathology of PD and Dementia by exploring high-quality preprocessed neuroimaging, replicating previous examinations of the data, and examining a wide variety of hypotheses without the impediment of the aforementioned barriers.

| Data Source In Nigeria | Clinical groups | Sex assigned at birth (Female or Male) | Number of participants |
|---|---|---|---|
| South-South region | Control | 11F, 25M | 36 |
| | PD | 7F, 8M | 15 |
| | Dementia | 10F, 9M | 19 |
| North Central region | Control | Nil | Nil |
| | PD | 1F, 4M | 5 |
| | Dementia | 7F, 6M | 13 |

Table 1: Data source from two geopolitical regions in Nigeria

## Methods

**Ethics**

The data sets were collected in Nigeria from three Imaging centers in two geopolitical regions of the country namely, the South-south region and the North-central region of Nigeria. Ethical Approval was obtained from the Research Ethics Committee of the University Of Port-Harcourt, Nigeria with the reference code: UPH/CERMAD/REC/MM84/056.

**Data sources:** Data is publicly available at <URL will be deposited at publication>.

**Neuroimaging data sources:** Data were collected from three Neuroimaging Diagnostic centers in Nigeria. Two centers are located in the South-South city of Port Harcourt and one in the Federal Capital Territory (FCT), Abuja. Data collection was approved by the University of Port Harcourt Research Ethics Review Committee and the ethics review boards of the diagnostic centers. The identified legal basis for processing this data for research is a public task as well as for the need to advance scientific research.

**Study participants**: Participants or their guardians signed informed consent forms for data collection for diagnostic and research purposes. A total of 261 brain MRI datasets were collected from 88 participants (Dementia: 32; Parkinson: 20; Control: 36), in Nigeria. The subjects were all Nigerians residing in Nigeria. Due to the sensitive nature of the datasets and to protect the confidentiality and privacy of data subjects, data protection measures such as pseudonymization (manual brainmask creation and masking), dedicated access control procedures and Data Use Agreement (DUA) were developed and utilized. There were 32 participants with Dementia (age: M = 65.10 years, SD = 14.49, Range = [36-86 ], 16 F, 16 M). There were 20 participants with PD (age: M = 62.7 years, SD = 13.20, Range = [41- 79], 8 F, 12 M). There were 36 healthy participants referred to here as controls (age: M = 32.92 years, SD = 10.19, Range = [10-56], 11 F, 25 M). Participants gave informed written consent that was approved by the Ethics Committee Board of the Lifebridge Diagnostic Lab, RUSTH lab and the Intercontinental Lab respectively. See **Table 1** for a breakdown of the demographics of the participant groups.

| Clinical group | Sex | Number of subjects | Age range (years) | Mean (years) | Standard deviation (years) |
|---|---|---|---|---|---|
| Control | M | 25 | 10-56 | 32.92 | 13.20 |
| | F | 11 | | | |
| PD | M | 12 | 41-79 | 62.7 | 13.2 |
| | F | 8 | | | |
| Dementia | M | 16 | 36-86 | 65.10 | 14.49 |

|   | F | 16 |   |   |   |

Table 2. **Demographics of participants.**

**Neuroimaging parameters**. Participants from the South-South lab 1 were imaged using a G Model 1.5-Tesla scanner. Participants from FCT lab were imaged using Model MRT-1503 scanner while participants from the South-South Lab 2 were imaged using aBRIVO MR235 0.3T Scanner. A 12-channel head coil was used at all sites.

For a subset of participants (19 total participants), multiple runs of data collection were performed (i.e. run-1, run-2). For an even smaller subset (17 total participants), contrast-enhanced T1w images were collected. 94% of those participants (16 participants) had ce-gadolinium, while the other had ce-deulonium as the contrast agent.

Due to the uniqueness of the data collection efforts, not all participants have equal numbers of images collected for each orientation, contrast enhancement, or run. Specifically, only 10 participants had 3 full orientations (axial, coronal, sagittal) of T1w images to use, while some had 0. This was also the case for T2w images, where only 3 participants had three orientations collected. For most participants where a FLAIR image was collected (36 participants total), only 9 participants had more than 1 orientation collected but never all 3. Information regarding the breakdowns of the number of images collected can be found in **Table 3.**

| Modality | Contrast-Enhanced | Run | Acquisition Orientation | Count |
|---|---|---|---|---|
| T1w | None | No run | axial | 13 |
| | | | coronal | 17 |
| | | | sagittal | 12 |
| | ce-gadolinium | No run | axial | 11 |
| | | | coronal | 9 |
| | | | sagittal | 12 |
| | ce-deulomin | No run | axial | 0 |
| | | | coronal | 0 |
| | | | sagittal | 1 |
| | None | Run 1 | axial | 3 |
| | | | coronal | 6 |
| | | | sagittal | 11 |
| | ce-gadolinium | Run 1 | axial | 1 |
| | | | coronal | 3 |
| | | | sagittal | 1 |
| | ce-deulomin | Run 1 | axial | 0 |
| | | | coronal | 0 |
| | | | sagittal | 0 |
| | None | Run 2 | axial | 3 |
| | | | coronal | 6 |
| | | | sagittal | 11 |
| | ce-gadolinium | Run 2 | axial | 1 |
| | | | coronal | 3 |
| | | | sagittal | 1 |
| | ce-deulomin | Run 2 | axial | 0 |
| | | | coronal | 0 |
| | | | sagittal | 0 |
| T2w | None | No run | axial | 25 |
| | | | coronal | 18 |
| | | | sagittal | 23 |
| | | Run 1 | axial | 5 |
| | | | coronal | 5 |
| | | | sagittal | 2 |
| | | Run 2 | axial | 5 |
| | | | coronal | 5 |
| | | | sagittal | 2 |
| FLAIR | None | No run | axial | 32 |
| | | | coronal | 12 |
| | | | sagittal | 0 |
| | | Run 1 | axial | 1 |
| | | | coronal | 0 |
| | | | sagittal | 0 |
| | | Run 2 | axial | 1 |
| | | | coronal | 0 |
| | | | sagittal | 0 |

Table 3. Number of images per subject, broken down by modality, contrast-type, run, and acquisition orientation.

**Anatomical data (T1w, T2w, FLAIR) preprocessing:** The raw data contains anatomical (T1 weighted, T2 weighted and FLAIR) magnetic resonance imaging data. The DICOMs were converted to BIDS using ezBIDS. Upon conversion, raw images were visually inspected using FSL's[15–17] *slicer* functionality implemented as brainlife.app.300, brainlife.app.301, and brainlife.app.689. Following this, quality metrics of the raw T1w and T2w images were computed using MRIQC[18] as brainlife.app.701 and brainlife.app.702.

Before publication, all images were masked to remove non-brain material in an effort to preserve participant anonymity. This was performed by generating brainmasks using FSL's *bet* tool implemented as brainlife Apps brainlife.app.2, brainlife.app.156, and brainlife.app.728. Masks were then manually refined within *fsleyes* due to imperfections in the generated brain masks. Following manual editing, the masks were applied to the anatomical data to extract brain data using FSL functionality implemented as brainlife.app.751, brainlife.app.752, and brainlife.app.753. Only defaced, masked images will be released to preserve participant anonymity.

**Table 4** lists all of the brainlife.io Apps, with their associated GitHub repositories and branches, used to process data.

| Application | Github repository | Open Service DOI | Git branch |
|---|---|---|---|
| MRIQC T1w | https://github.com/brainlife/app-mriqc | https://doi.org/10.25663/brainlife.app.701 | 22.0.6 |
| MRIQC on T2w | https://github.com/brainlife/app-mriqc | https://doi.org/10.25663/brainlife.app.702 | 22.0.6 |
| FSL Brain Extraction (BET) on T1w | https://github.com/brainlife/app-FSLBET | https://doi.org/10.25663/bl.app.2 | 2.0 |
| FSL Brain Extraction (BET) on T2w | https://github.com/brainlife/app-FSLBET | https://doi.org/10.25663/brainlife.app.156 | 2.0 |
| FSL Brain Extraction (BET) on FLAIR | https://github.com/brainlife/app-FSLBET | https://doi.org/10.25663/brainlife.app.728 | 2.0 |
| Apply mask to extract brain data - T1w | https://github.com/brainlife/app-FSLBET | https://doi.org/10.25663/brainlife.app.751 | apply-mask-v1.0 |
| Apply mask to extract brain data - T2w | https://github.com/brainlife/app-FSLBET | https://doi.org/10.25663/brainlife.app.752 | apply-mask-v1.0 |
| Apply mask to extract brain data - FLAIR | https://github.com/brainlife/app-FSLBET | https://doi.org/10.25663/brainlife.app.753 | apply-mask-v1.0 |
| T1w images for figures | https://github.com/brainlife/app-slicer-fsl | https://doi.org/10.25663/brainlife.app.300 | master-app-v1.0.0 |
| T2w images for figures | https://github.com/brainlife/app-slicer-fsl | https://doi.org/10.25663/brainlife.app.301 | master-app-v1.0.0 |
| FLAIR images for figures | https://github.com/brainlife/app-slicer-fsl | https://doi.org/10.25663/brainlife.app.689 | master-app-v1.0.0 |

**Table 4.** Description and web links to the open-source code and open cloud services used in the creation of this dataset.

## Data Records

**T1-weighted Anatomical.** T1w images were collected as 2D images, with one high-resolution and two low-resolution planes. For some participants, multiple scans were collected in a single session and designated by a "run" tag, with the first run being "run-1" and the second being "run-2". For the others who did not have multiple runs collected, there will be a "no-run" tag. Some data was collected with contrast agents, specifically ce-gadolinium and ce-deulomin. These images are tagged with the appropriate contrast agent.

```
upload/sub-{}/anat/
      sub-{}_acq-[coronal,sagittal,axial]_tag-brainextracted_tag_desc-{}_T1w.json
      sub-{}_acq-[coronal,sagittal,axial]_tag-brainextracted_tag_desc-{}_T1w.nii.gz
```

**T2-weighted Anatomical.** T2w images were collected as 2D images with one high-resolution and two low-resolution planes. For some participants, multiple runs were also collected, and were designated by a "run-" tag, with the first run being "run-1" and the second being "run-2". For the others who did not have multiple runs collected, there will be no "run-" tag.

```
upload/sub-{}/anat/
       sub-{}_acq-[coronal,sagittal,axial]_tag-brainextracted_tag_desc-{}_T2w.json
       sub-{}_acq-[coronal,sagittal,axial]_tag-brainextracted_tag_desc-{}_T2w.nii.gz
```

**Fluid Attenuated Inversion Recovery (FLAIR).** FLAIR images were collected as 2D images with one high-resolution and two low-resolution planes. For some participants, multiple runs were collected, and were designated by a "run" tag, with the first run being "run-1" and the second being "run-2". For the others who did not have multiple runs collected, there will be a "no-run" tag.

```
upload/sub-{}/anat/
       sub-{}_acq-[coronal,sagittal,axial]_tag-brainextracted_tag_desc-{}_FLAIR.json
       sub-{}_acq-[coronal,sagittal,axial]_tag-brainextracted_tag_desc-{}_FLAIR.nii.gz
```

**Brain masks.** Brainmasks were generated for each individually collected image. First-pass brainmasks were generated using FSL's *bet* functionality, and then downloaded locally in order to be manually refined. Upon refinement, data were then reuploaded to brainlife.io before being applied to the anatomical images.

```
upload/sub-{}/mask/
       Sub-{}_acq-[coronal,sagittal,axial]_tag-anat_tag-brain_tag-[t1,t2,flair]_tag_desc-{}_mask.nii.gz
```

**MRIQC.** MRIQC was used to compute quality assurance metrics of the raw T1w and T2w images. With this comes a data structure that loosely follows the BIDS standard.

```
regressors/
       regressors.tsv
```

## Technical Validation

**Anatomical (T1w, T2w, FLAIR) raw data.**
In this section, we provide a qualitative and quantitative evaluation of the data derivatives made available on brainlife.io. **Figure 1** describes the workflow used to process and publish the data. Raw DICOM files from the MRI scanners were first converted to BIDS standard formats using ezBIDS and uploaded directly to brainlife.io.

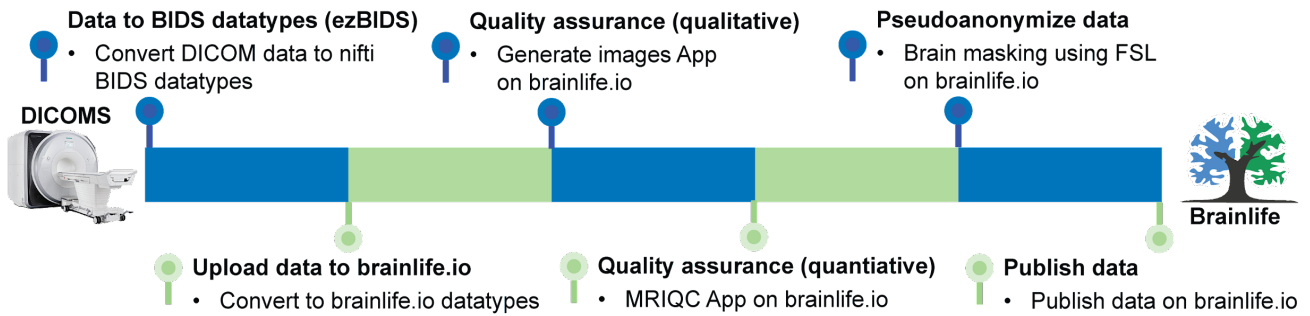

**Figure 1. Workflow for releasing the Nigerian Brain Dataset.** Diagram describing the workflow performed for getting the Nigerian Brain Dataset ready for publication. Raw DICOMs from the MRI scanners were first converted into BIDS standard

format using ezBIDS, and the converted data was uploaded and organized automatically to brainlife.io. Data was then assessed for quality both qualitatively using brainlife.io Apps to generate images of the raw data, and quantitatively using MRIQC. Before release, the data was pseudo-anonymized via defacing by creating brainmasks and removing all non-brain material. Defaced data was then published with a digital object-identifier (DOI) which can be interacted with via brainlife.io.

Upon conversion of the data from DICOMS to BIDS-formatted data types, data were masked and were visually inspected for quality. **Figure 2** exemplifies the quality of the raw anatomical [T1w & T2w (**a**), FLAIR (**b**)] images obtained with [brainlife.app.300](brainlife.app.300), [brainlife.app.301](brainlife.app.301), and [brainlife.app.689](brainlife.app.689) in representative participants from each diagnosis-group [i.e. Dementia (sagittal T1w & T2w), Parkinson's (coronal T1w & T2w), Control (axial T1w & T2w)]. These images are representing the highest resolution plane of the collected scan for each modality. Note, for FLAIR images, no high-resolution sagittal plane was collected.

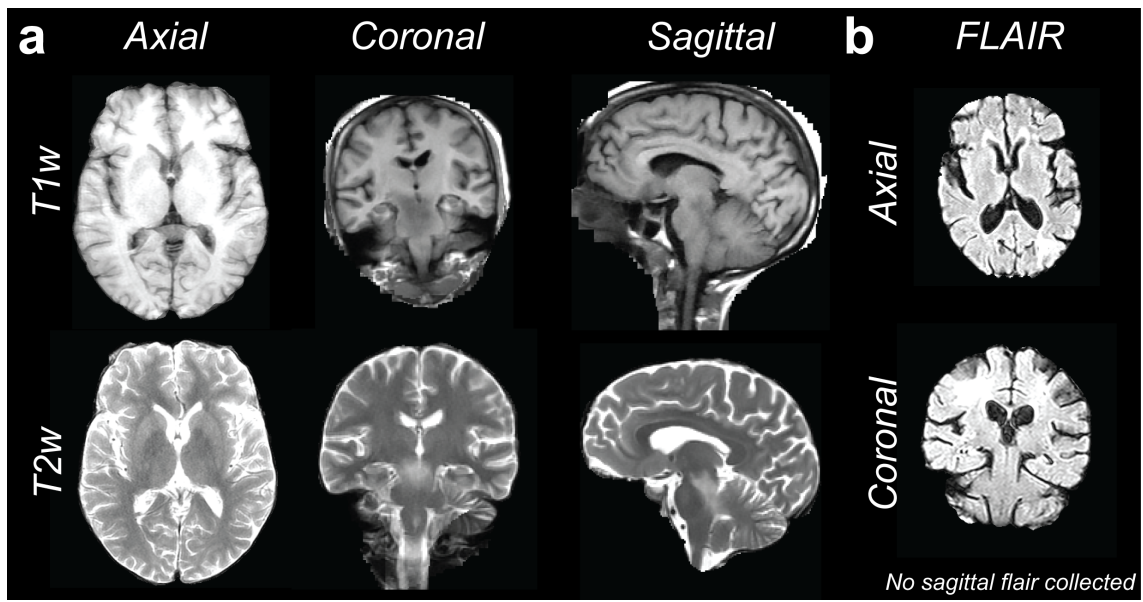

**Figure 2. Images of raw anatomical (T1w, T2w, FLAIR) data. a.** Axial (*left*), coronal (*middle*), and sagittal (*right*) high-resolution planes for the raw T1w and T2w anatomical data collected. Shown axial image is from a control participant, coronal image is from a Parkinson's patient, and the sagittal image is from a Dementia participant. **b.** Axial (*top*) and coronal (*bottom*) high-resolution planes for the raw FLAIR images collected from a control participant.

There exists a wide range of quality of the collected data. To assess this, following visual quality assurance, MRIQC[18] was performed on the raw T1w and T2w images for each high-resolution plane collected. **Figure 3** reports the contrast-to-noise ratio (CNR) calculated by MRIQC for each participant, modality, and image plane.

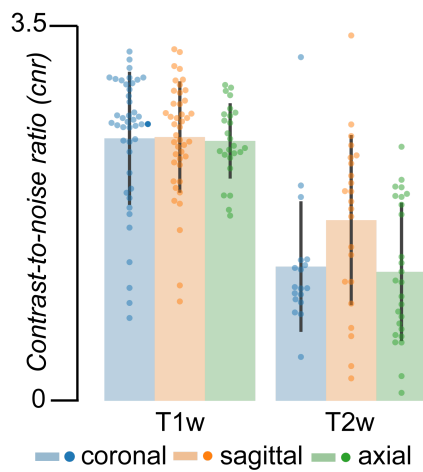

**Figure 3. Range of quality of data (MRIQC data) and the failure rate for running MRIQC. a.** Contrast-to-noise ratio across all of the T1w and T2w data for each high-resolution plane (coronal: *blue*, sagittal: *orange*, axial: *green*). Error bars represent ± 1 standard deviation.

## Usage Notes

**Data Access Conditions/Requirements**
Access conditions/requirements are considered necessary due to the nature of the datasets - personal data (sensitive health data) - to protect the rights of the participants. The datasets can be classified as special category data and as such requires technical and organizational measures or safeguards to ensure the confidentiality and privacy of the participants[20,21]. Rather than limit access to the datasets, such measures are designed to enable responsible research and innovation (RRI). To achieve this, data protection measures such as pseudonymization, access control and Data Use Agreement (DUA) have been developed [22]. Users are requested to agree to the DUA before accessing the data. All shared datasets were pseudonymised using masks generated using FSL[15–17] that were then manually edited to remove all non-brain material that could be useful in re-identification procedures.

Access to the datasets is granted on a 'controlled access' basis through an access review committee. Users will be fully identified and technically authenticated after providing information on their purpose of use which will be evaluated by the access review committee (ARC). All users will be asked to sign a DUA detailing their responsibilities and liabilities while the data is in their possession. This is a legally binding agreement that users are expected to comply with in their role as data controllers of the data they process. The DUA would be signed prior to providing access to the data.

**Accessing the data**
The data are available on brainlife.io and can be found using the following DOI: https://doi.org/10.25663/brainlife.pub.45. The following video shows how to access, download and visualize the data published with the record: https://www.youtube.com/watch?v=QEWEQydpbz4.

## Data Use Agreement

A data use agreement is provided as a requirement for accessing the data directly on brainlife.io. Datasets are made freely available to anyone that agrees to the terms.

Data files can also be downloaded, and some can be organized into BIDS standard[23]. The data derivatives are stored in numerous formats, including NIFTI, tab-separated values (tsv), html, and text files. Access to the published data is currently supported via (i) web interface and (ii) Command Line Interface (CLI).

The brainlife.io CLI can be installed on most Unix/Linux systems using the following command:

```
npm install brainlife.io -g
```

The CLI can be used to query and download partial or full datasets. The following example shows the CLI command to download all T1w datasets from a subject in the publication data:

```
bl pub query # this will return the publication IDs
bl bids download --pub 64233a2a73d2685502db46a3 --subject 01 --datatype \
neuro/anat/t1w --tag "acq-coronal"
```

The following command downloads the data in the entire project (from Release 1) into BIDS format:

```
bl bids download --pub 64233a2a73d2685502db46a3
```

Additional information about the brainlife.io CLI commands can be found at https://github.com/brainlife/cli.

## Code availability

**Table 4** reports the links to each web service and github.com URL implementing the processing pipeline. Analyses using MRIQC outputs were performed using open source code hosted by brainlife.io and available as a Jupyter Notebook at
https://github.com/bacaron/nigerian_brain_analyses/blob/main/nigerian_brain_analyses.ipynb.

## Acknowledgements

National Science Foundation (NSF) awards 1916518, 1912270, 1636893, and 1734853. National Institutes of Health awards (NIH) R01MH126699, R01EB030896, R01EB029272 and a Microsoft Investigator Fellowship to Franco Pestilli. A Wellcome Trust award (226486/Z/22/Z) and a gift from the Kavli Foundation to Franco Pestilli and Damian Eke.

## Author Contributions

| | |
|---|---|
| Eberechi Wogu | - Data collection and conceptualization, writing |
| Patrick Filima | - Data collection and preparation of data |
| Tawe Godwin | - Data collection |
| Damian Eke | - Conceptualization, writing |
| Simi Akintoye | - Conceptualization |
| George Ogoh | - Conceptualization |
| Catherine Leal | - Data processing |
| Mohammed F. Mehboob | - Data processing |

Dan Levitas, Bradley Caron, Peer Herholz, Soichi Hayashi and Franco Pestilli - Software development, data processing, curation, writing, and study conceptualization

## Competing Interests

The authors declare no competing interests.